\documentclass{article}
\usepackage[dvips]{graphicx}
\begin{document}
\title{Elimination of spiral chaos by periodic force 
for the Aliev-Panfilov model}
\author{Hidetsugu Sakaguchi and Takefumi Fujimoto\\
Department of Applied Science for Electronics and Materials,\\ Interdisciplinary Graduate School of Engineering Sciences,\\
 Kyushu University, Kasuga, Fukuoka 816-8580, Japan}
\maketitle
\begin{abstract}
Spiral chaos appears in the two dimensional Aliev-Panfilov model. 
The generation mechanism of the spiral chaos is related to the breathing instability of pulse trains. The spiral chaos can be eliminated by applying periodic force uniformly.  The elimination of spiral chaos is most effective, when the frequency of the periodic force is close to that of the breathing motion. 
\end{abstract}
\bigskip
\bigskip
\bigskip

Some types of cardiac arrhythmia are characterized by rotating waves, which are similar to 
spiral waves found in excitable media \cite{rf:1}. Control and elimination of arrhythmia and the spiral waves are medically important.  
The control of regular spiral patterns in excitable media has been studied with several methods. Meandering of the spiral core can be controlled by a periodic parameter modulation \cite{rf:2}, impulses and periodic force have been applied to suppress spiral waves \cite{rf:3,rf:4}, and local and global feedback have been applied to excitable systems to eliminate spiral waves \cite{rf:5}.   A serious cardiac arrhythmia such as ventricular fibrillation is related to spiral chaos where many spiral cores are spontaneously generated. 
To eliminate spiral chaos is more important. 
   
Periodic forcing to spatio-temporal chaos was used to restore regular waves 
for diffusively coupled chemical oscillators and the complex Ginzburg-Landau 
equation \cite{rf:6,rf:7}. In this paper, we attempt to control and eliminate numerically spiral chaos by applying 
 a periodic force and find an effective frequency for the elimination.
We use the Aliev-Panfilov model for the cardiac cell \cite{rf:8}:
\begin{eqnarray} 
\frac{\partial e}{\partial t}&=&-K(e-a)(e-1)-er+\nabla^2 e,\nonumber\\
\frac{\partial r}{\partial t}&=&[\epsilon+\mu_1r/(\mu_2+e)][-r-ke(e-b-1)].
\end{eqnarray}
Here $e$ stands for the membrane potential and $r$ stands for the conductance of the inward current. 
This model is a phenomenological model which represents certain feature of impulse propagation in cardic tissue.  The parameter values of $K,\,a,\,\epsilon,\,\mu_1,\,\mu_2$ and $b$ are evaluated based on the real experiment.  The model equation exhibits spiral breakup and spiral chaos in a certain parameter range \cite{rf:9}. 

\begin{figure}[htb]
\begin{center}
\includegraphics[width=12cm]{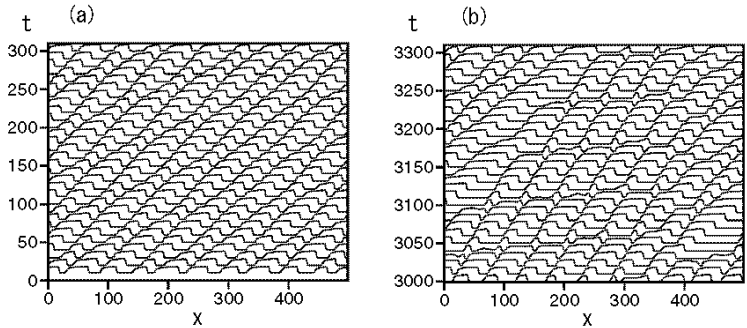}
\caption{(a) Time evolution of the breathing state at $a=0.063$ for the pulse train with wavenumber $k=2\pi\cdot 10/L\sim0.126$.  (b) Time evolution of the 
wavenumber decrease process at $a=0.044$.}
\label{fig:1} 
\end{center}
\end{figure}

\begin{figure}[htb]
\begin{center}
\includegraphics[width=8cm]{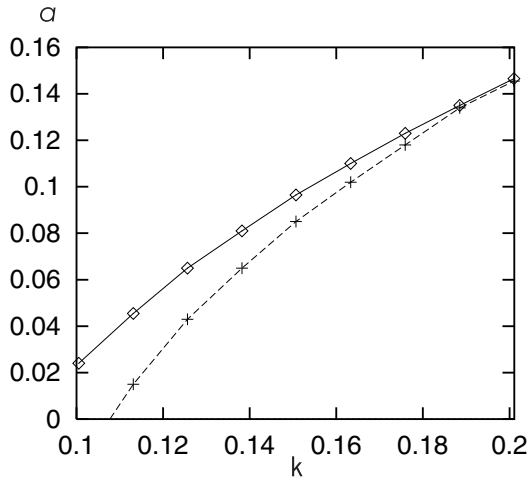}
\caption{Bifurcation curves as a function of wavenumber $k$ of the pulse trains. The solid curve denotes the breathing instability, and the dashed curve denotes the bifurcation below which the wavenumber decreasing transition occurs.}
\label{fig:2} 
\end{center}
\end{figure}

\begin{figure}[htb]
\begin{center}
\includegraphics[width=12cm]{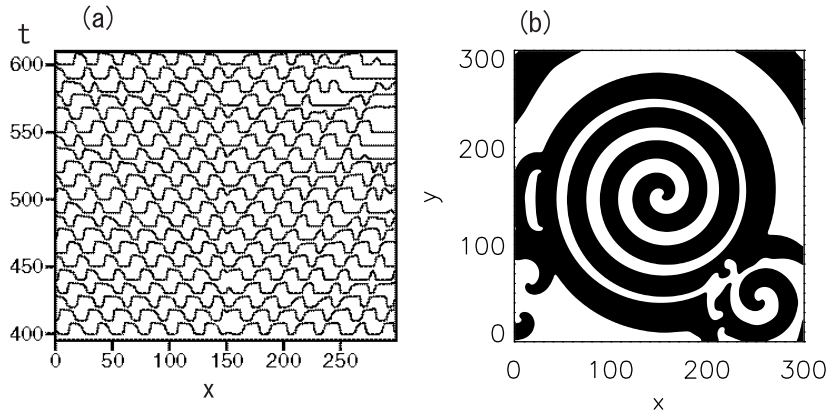}
\caption{Time evolution of $e(x,L-x)$ for $a=0.115$.
The pulse trains emitted form the spiral core exhibit breathing motion and 
the pulse collapse occurs at $t\sim 520$. (b) Snapshot of $e$ at $t=800$. In the shaded region, $e$ is larger than 0.4.}
\label{fig:3} 
\end{center}
\end{figure}

\begin{figure}[htb]
\begin{center}
\includegraphics[width=12cm]{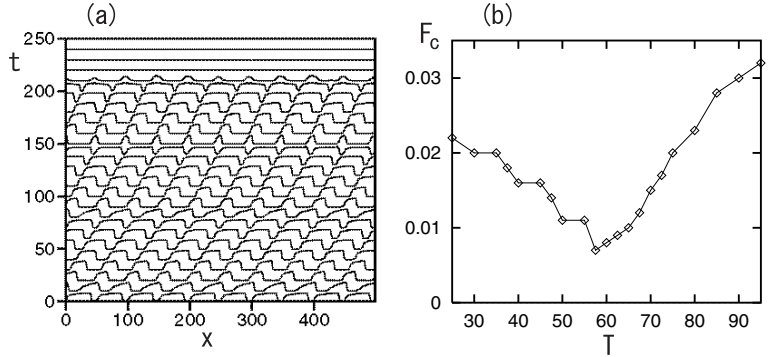}
\caption{(a) Time evolution for $a=0.06$, $k=2\pi\cdot 10/L,\,F=0.009$ and $\omega=2\pi/60$. For $t>220$, the pulse train is completely 
collapsed. (b) Critical values of $F$ as a function of the period $T=2\pi/\omega$ of the external force for $a=0.06$.}
\label{fig:4} 
\end{center}
\end{figure}

\begin{figure}[htb]
\begin{center}
\includegraphics[width=13cm]{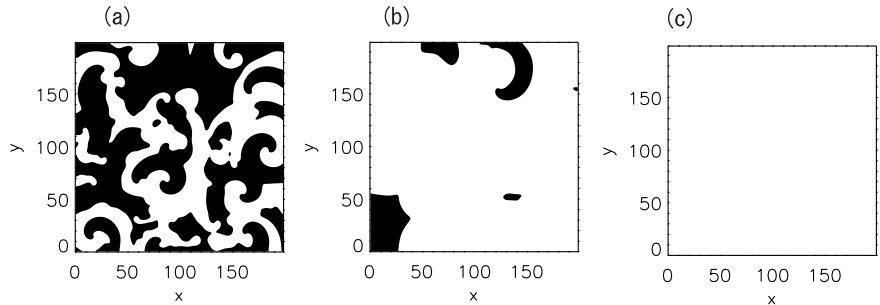}
\caption{Three snapshots of pattern of $e$ at $t=0,100$ and $300$ for $a=0.1, F=0.036$ and $\omega=2\pi/60$. The spiral chaos disappears completely at $t=300$.}\label{fig:5} 
\end{center}
\end{figure}

\begin{figure}[htb]
\begin{center}
\includegraphics[width=8cm]{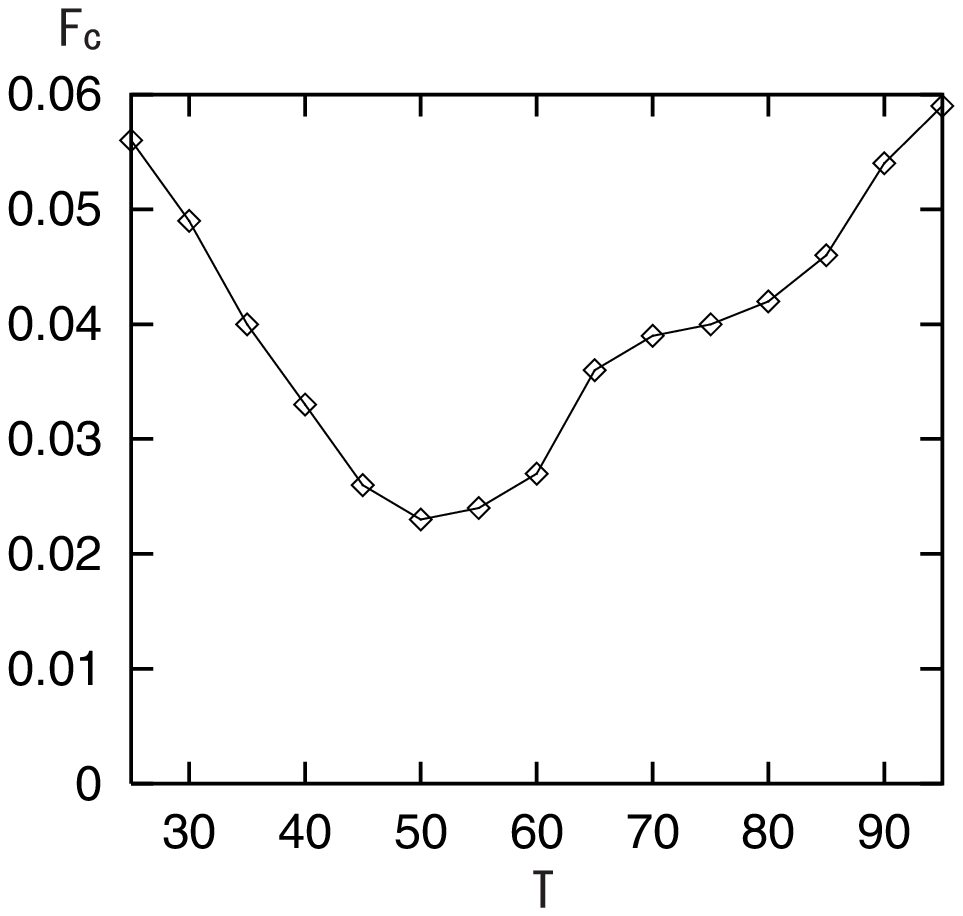}
\caption{Critical strength $F$ for the spiral collapse at $a=0.1$ as a function of the period $T$ of the external force.}
\label{fig:6} 
\end{center}
\end{figure}

We perform numerical simulations of Eq.~(1). The parameters except for $a$ 
are fixed as 
$K=8,\,\epsilon=0.01,\,\mu_1=0.11,\,\mu_2=0.3$ and $b=0.1$. 
Firstly, we show numerical results for a one-dimensional system, 
that is, $\nabla^2 e$ in Eq.~(1) is replaced by $\partial^2e/\partial x^2$. 
The numerical simulation was performed using the finite difference method with $\Delta t=0.005$ and $\Delta x=0.5$. 
The system size $L$ is 500, and the periodic boundary conditions are used.    
We have found that a pulse train propagates with a constant velocity for large $a$, however, the pulse train exhibits a breathing instability 
by decreasing $a$. 
Figure 1(a) displays the time evolution of $e(x)$ at $a=0.063$ for the pulse train with wavenumber $k=2\pi\cdot 10/L\sim 0.126$. 
The pulse width is temporally oscillating. The oscillatory instability of a single pulse in a one-dimensional ring of excitable media is called the alternans instability, since the pulse width almost alternates  every time when the pulse turns around the ring \cite{rf:10}. 
The oscillatory instability of a pulse train with spatial period $2\pi/k$ almost corresponds to the alternans instability in a ring of size $2\pi/k$. 
The alternans instability in a ring of size $2\pi/k=L/10$ occurs as a supercritical Hopf bifurcation at $a=a_{c0}\sim 0.0647$. 
However, the oscillatory instability of the pulse train occurs at slightly larger $a=a_{c}\sim 0.0664$ for $k=2\pi\cdot 10/L$, accompanying spatial modulation. As seen in Fig.~1(a), the phase of the breathing motion is not synchronized for all pulses in the pulse train, as the result of the spatial modulation.  
As $a$ is decreased further, the breathing amplitude becomes larger, and the spatial modulation grows. Finally, one pulse is annihilated and the 
wavenumber is decreased to $k=2\pi\cdot 9/L$.  The time evolution of the 
wavenumber decrease process is shown in Fig.~1(b) for $a=0.044$.
It is characteristic of the cardiac tissue that the pulse width is easily varied  with spatial periods of pulse trains. This characteristics is involved in the  Aliev-Panfilov model. 
The critical parameters for the breathing instability and the wavenumber changing bifurcation strongly depend on the pulse interval. 
Figure 2 displays the two bifurcation curves (the breathing instability and the wavenumber decreasing bifurcation) as a function of wavenumber $k$. 
As the pulse interval $2\pi/k$ is narrower, the instabilities occur more easily, and the two bifurcation curves approach each other. 
The wavenumber changing process occurs just after the breathing instability for $k\sim 0.2$. 

Next, we consider the instability of the spiral pattern in two dimensions. 
A stationary rotating spiral becomes unstable in a certain parameter range as shown by Panfilov \cite{rf:9}.
We have performed a numerical simulation of Eq.~(1) in two dimensions. The system size is $300\times 300$ and no flux boundary conditions are used. A spiral pattern is stable for $a>0.13$ for the fixed parameters $K=8,\,\epsilon=0.01,\,\mu_1=0.11,\,\mu_2=0.3$ and $b=0.1$. The initial condition is a regular spiral pattern obtained numerically for $a=0.13$.   
Figure 3(a) displays a time evolution of $e(x,y)$ on the line $y=L-x$ at $a=0.115$.
At this parameter, pulse trains with wavenumber $k\sim 0.185$ are emitted from the spiral core. 
A one-dimensional pulse train with wavenumber $k\sim 0.185$ is unstable for the breathing motion at $a=0.115$ as is shown in Fig.~2.
The pulse trains indeed exhibit breathing motion, and the wavenumber decreasing processes occur at $t\sim 520$. 
In one-dimensional system, the wavenumber decreasing process leads to a more stable structure with smaller wavenumber, however, in two dimensions, the wavenumber decreasing process leads to formation of additional topological defects, the spiral breaks up, and then spiral chaos appears. This is another route of spiral breakup, although it is similar to the spiral breakup via a wavenumber changing process by the Eckhaus instability \cite{rf:11}.  
Figure 4(b) displays a snapshot of $e$ at $t=800$, where $e$ takes a larger value than 0.4 in the shaded region.  Two main spirals and several small spirals are generated as a result of the spiral breakup.

To eliminate the pulse train and the spiral chaos, we apply an external periodic force.
The model equation is written as 
\begin{eqnarray} 
\frac{\partial e}{\partial t}&=&-k(e-a)(e-1)-er+\nabla^2 e+F\sin(\omega t),\nonumber\\
\frac{\partial r}{\partial t}&=&[\epsilon+\mu_1r/(\mu_2+e)][-r-ke(e-b-1)],
\end{eqnarray}
where $F$ and $\omega$ are the amplitude and frequency of the external periodic force.  We first show a numerical result of a one-dimensional system. 
Figure 4(a) displays a time evolution for $F=0.009$, $a=0.06$, $k=2\pi\cdot 10/L$ and $\omega=2\pi/60$. The initial condition is a breathing state for Eq.~(1)
 without the external force. 
The breathing motion of the pulse width is entrained to the external force. 
The amplitude of the breathing motion grows, and finally  
the pulse train structure collapses completely for $t>230$.  Thus, the pulse train could be eliminated by applying the external periodic force. 
We have investigated a critical value of the amplitude $F$ for the complete collapse. Figure 4(b) displays the critical values $F_c$ as a function of the period $T=2\pi/\omega$ for $a=0.06$.   
The phase diagram has a shape like the Arnold tongue for the forced entrainment, and the critical value $F_c$ takes the smallest value at $T\sim 57.5$. 
The period is close to the period $T^{\prime}\sim 61$ of the natural breathing motion without the periodic forcing.
 These results are interpreted as a kind of resonance. That is, if the period of the external force is close to the period of the natural breathing motion, the effect of the external force is enhanced and the pulse train collapses easily. 

We have applied a periodic force to eliminate the spiral chaos in two dimensions.
 The system size is $200\times200$ and the parameter $a$ is 0.1. The spiral chaos appears for this parameter.  We have used a snapshot of the spiral chaos as an initial condition for the forced system. 
In the simulation of the two dimensional system, we have applied 
the periodic force $F\sin\omega t$ to the spiral chaos only for $0<t<6\pi/\omega$ (three periods), and observed whether the spiral chaos collapses or the spiral chaos is maintained after the periods of the external forcing. 
The three periods are sufficient to see the effect of the external forcing. 
Figure 5 displays three snapshots of pattern of $e$ at $t=0,100$ and $300$ for $a=0.1, F=0.036$ and $\omega=2\pi/60$. The spiral chaos collapses completely and the uniform state $e(x,y)=0$ and $r(x,y)=0$ has been obtained after the application of external force.    
We have numerically obtained the critical value $F_c$ for the collapse of the spiral chaos. The critical strength $F_c$ is shown in Fig.~6 as a function of the period $T=2\pi/\omega$.  The critical curve takes minimum at $T\sim 50$.  It is close to the period $\sim 55$ of the natural breathing motion at this parameter.    

To summarize, we have performed numerical simulations of the Aliev-Panfilov model.
We have found a breathing instability for the pulse trains in one dimension.  The breathing instability leads to spiral breakup in two dimensions.  
The periodic force is applied to eliminate the pulse trains and the spiral chaos. We have found that the most effective frequency to eliminate the wave patterns is close to the natural frequency of the breathing motion. This is interpreted as a kind of the resonance effect.  The periodic forcing with the most effective frequency may be effective as a method of mild defibrillation.

\end{document}